***FREQUENCIES OF FLARE OCCURRENCE: INTERACTION BETWEEN CONVECTION AND CORONAL LOOPS***

Short title: Slopes of flare energy spectra

D. J. Mullan[1] and R. R. Paudel[1]

[1]Department of Physics and Astronomy, University of Delaware, Newark DE 19716

Corresponding author: mullan@udel.edu

*Abstract*

Observations of solar and stellar flares have revealed the presence of power law dependences between the flare energy and the time interval between flares. Various models have been proposed to explain these dependences, and to explain the numerical value of the power law indices. Here, we propose a model in which convective flows in granules force the foot-points of coronal magnetic loops, which are frozen-in to photospheric gas, to undergo a random walk. In certain conditions, this can lead to a twist in the loop, which drives the loop unstable if the twist exceeds a critical value. The possibility that a solar flare is caused by such a twist-induced instability in a loop has been in the literature for decades. Here, we quantify the process in an approximate way with a view to replicating the power-law index. We find that, for relatively small flares, the random walk twisting model leads to a rather steep power law slope which agrees very well with the index derived from a sample of 56,000+ solar X-ray flares reported by the GOES satellites. For relatively large flares, we find that the slope of the power law is shallower. The empirical power law slopes reported for flare stars also have a range which overlaps with the slopes obtained here. We suggest that in the coolest stars, a significant change in slope should occur when the frozen-flux assumption breaks down due to low electrical conductivity.

1. **Introduction**

Flares on stars are observed to have a range of energies. Shakhovskaya (1989: hereafter Sh89) reported on a study of flares from 25 different stars, using a filter to isolate one particular band of wavelengths in the optical. Sh89 found that in this band, the flare energy ranged from $10^{27}$ ergs to $10^{35.5}$ ergs, i.e. a range of 8-9 orders of magnitude in energy. Sh89 plotted the flare energy versus the frequency of flare occurrence f for each of star in her sample: the value of f at energy E was chosen by Sh89 to be the frequency of flares "with energy exceeding E". We refer to such a plot as an "integral energy spectrum".

Examples of such spectra are illustrated in Figure 1. Lines in the plot with different symbols refer to different objects: the data have been excerpted from Fig. 38 of Gershberg (2002: hereafter G02). The data refer to flares on the Sun, to flares on 6 solar neighborhood flare stars, and to flares on two stars in young clusters (Orion, Pleiades). In quantitative terms, the flare energies in Fig. 1 range from $\approx 10^{26}$ ergs (for a solar flare) to $\approx 10^{36}$ ergs (for a flare star in the Orion cluster). Also superposed in Fig. 1 are examples of two power laws (with arbitrary zero points) which will be derived in the present paper (see eqs. 10 and 11). For future reference, we note that the steeper power law (index = 1.5) fits most of the energy spectrum for the Sun, while the shallower power law (index = 1.0) would fit the Orion flare spectrum well if a slight shift were applied to the zero-point.

The most important characteristic of the "energy spectra" in Fig. 1 is as follows: for every star in the sample, flares with lower energy occur more frequently, while flares with larger energies occur more rarely. More quantitatively, the curve for each star in Fig. 1 is found to exhibit monotonic behavior which can be represented well by three regimes, as follows.

(i) For the largest flares in certain stars, the spectrum becomes flatter as we approach the left-hand side of the figure, as if approaching a horizontal asymptote in energy: this behavior can be seen in Fig. 1 in the case of the Sun, and in the cases of the two solar neighborhood flare stars UV Cet and AD Leo. If the asymptotic energy is physically real (rather than an artifact of a scarcity of data for the rarest, and most energetic, flares), it could mean that there is a maximum energy which can be released in a flare on any particular star.

(ii) For the smallest flares, observational selection makes it harder to identify a "true" flare against the background of light from the star which is always subject to some fluctuations, either due to the Earth's atmosphere, or due to "micro-flares" in the star. As the observing conditions become noisier, the range of these "lost flares" extends to progressively higher energies. The "lost flares" were not included in G02's Fig.38, and as a result, we have not included them in our Fig. 1: if they were to be included, they would appear as a precipitous decline in each curves at the right-hand edge. Examples of such precipitous declines can be found in other figures in G02s (e.g. Figs. 36, 37, and 39), and also in Fig. 1 of Kasinsky and Sotnicova (2003). We shall refer further to regime (ii) in Sections 13 and 14 below.

(iii) At intermediate energies, a power-law with a negative slope (–b) can be fitted to the spectrum over a range which may (in some cases) span up to 1-2 orders of magnitude in f. Regime (iii) in solar neighborhood flare stars may extend from f ≈ 1 per hour to f ≈ 0.001 per hour. Thus, the time scales between flares in regime (iii) can be as short as a few hours, and as long as tens of days.

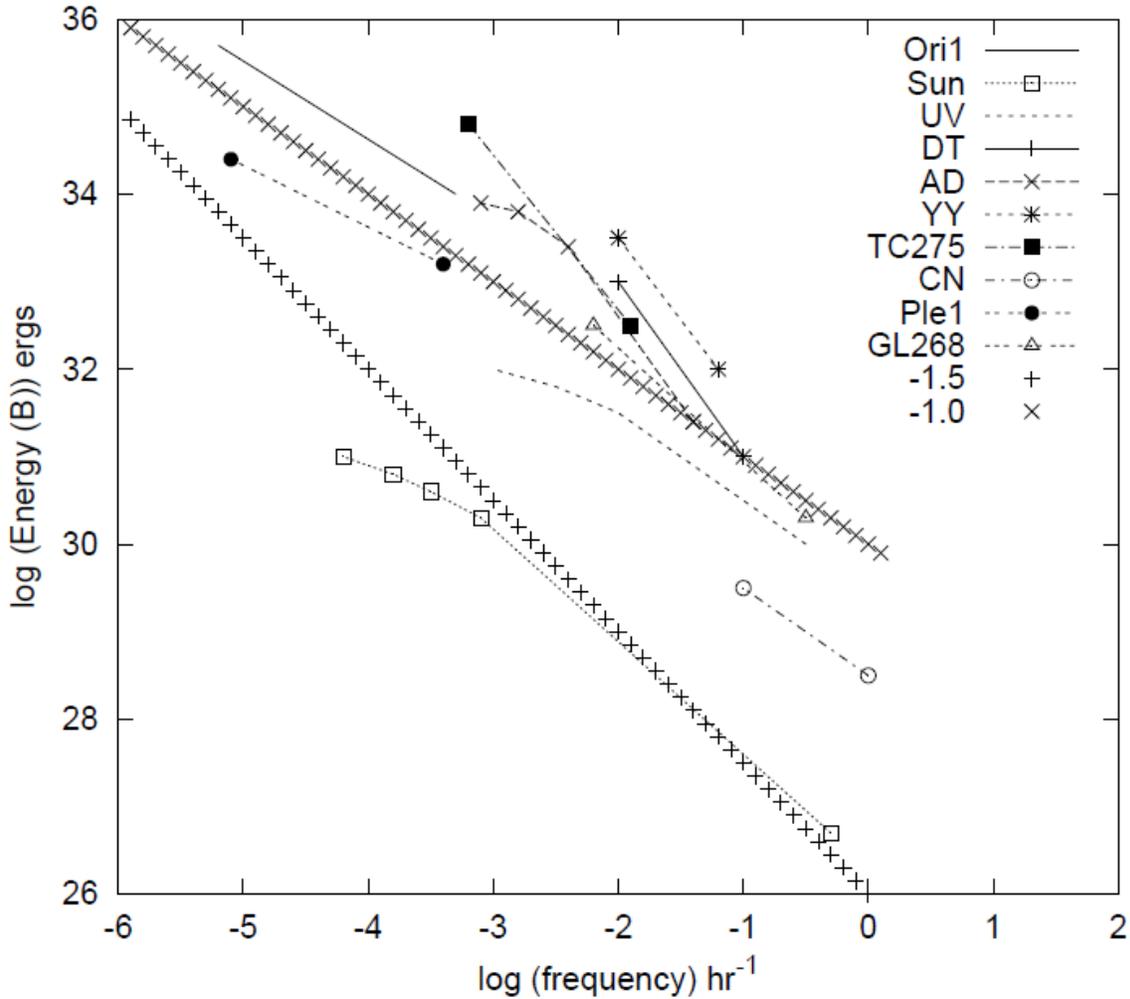

Figure 1. Flare energy spectra for solar flares and for flares on 9 stars (excerpted from G02, Fig. 38). The vertical axis indicates the energy (in units of ergs) which is emitted during a flare in the visual B band (extending between wavelengths of 360 and 550 nm). The frequency is in units of inverse hours. Two power laws are overplotted to guide the eye: the steeper power law has an index of -1.5, while the shallower has an index of -1.0. These are the power laws which we derive in the present paper for two categories of flares (see eqs. (11) and (10) respectively).

In terms of flares which can be confidently identified (i.e. regimes (i) and (iii)), we shall refer to the "standard flare energy spectrum" as one which exhibits a rather steep power-law spectrum at lower energies and a flatter spectrum at the largest energies. We refer to this standard spectrum as a "broken power law": further discussion of the broken power law will be presented in Section 15.2.

For future reference, it is important to note the convention that we have adopted in Fig. 1: we plot energy along the vertical axis and frequency along the horizontal axis: this is the convention chosen by Sh89 and by Gershberg. However, in other papers (e.g. Hawley et al 2014), the authors reverse the choice, plotting frequency along the vertical axis. Thus the behavior that we refer to as a "flattening" on the left-hand side of the spectrum in regime (i) appears in Hawley et al (2014) as a steepening on the right-hand side of the spectrum. And as regards regime (ii), whereas we speak of a steepening on the right-hand side of the spectrum, Hawley et al (2014) find a flattening on the left-hand side of the spectrum.

In Sections 2-4, we summarize the empirical power-laws which have been determined for flare energy spectra using optical and X-ray photons.

In this paper, our aim is to see if we can understand the slopes in regimes (iii) and (i) in terms of a magnetic loop model (summarized in Section 5). The essential aspect of our model is that, since the loop is embedded in the photosphere, where the foot-point of each field line may be "frozen in" to the highly conducting local gas, the field lines can be subjected to a random walk due to convective motions in the photospheric granules. This random walk can, in favorable conditions, lead to a twisting of the loop: a flare is expected to occur if the twist exceeds a certain critical value.

To set the stage for our calculations, we use solar data to describe the relevant properties of magnetic loops (Sections 6, 7) and of granules (Section 8). The basic physical process which we consider in this paper is described in Section 9. Power-law indices are derived in Sections 10 and 11. In order to ensure that our model is consistent with the observed energies of flares, we estimate the field strengths which are required in the Sun and in flare stars in order that magnetic energy can in fact provide the observed flare energies (Sections 12, 13). In the case of the coolest stars, the "frozen-in" condition is expected to break down: the effects of this are explored in Section 14. Discussion of certain aspects of flares is in Section 15. Conclusions are in Section 16.

## 2. Energy Spectra in Terms of the Flare <u>Frequency</u> f: Stars

Empirically, the integrated spectrum, i.e. the relationship between the flare energy E and the frequency of flares with energies ≥E, is typically (e.g. G02) represented as

log f(≥E) = a – b*log(E)            (1)

Equivalently, we note that the number dN of flares with energies in the range from E to E+dE, i.e. the differential energy spectrum, can be written as $dN/dE = E^{-\alpha}$. Comparing this with eq. (1), we see that the numerical value of α is equal to b+1.

### 2.1 Ground-based data

Ground-based observations have led (see G02) to determination of the constants a and b in a sample of 16 flare stars in the solar neighborhood and in young star clusters. The ratio of a/b for each star gives the value of log(E) for flares which occur hourly on that star: the a/b values range from $10^{23}$ ergs (V1216 Sgr, sp. Type dM4.5e) to $10^{31}$ ergs (BY Dra, dK6e). The largest flare energies are found among young stars in clusters (Orion, Pleiades), where flare energies up to $10^{35.5}$ ergs are observed, albeit at very low frequencies (once every $10^5$ hours). Clearly, different stars generate flares which differ in energy by many orders of magnitude.

As regards the parameter b, G02 lists values for 16 solar neighborhood stars. These range from 0.47±0.09 for DT Vir (dM1.5e) to 1.10±0.09 for CN Leo (M6.5e). The range of b values is much more limited than the broad range of energies (a/b) which emerge in once-per-hour flares. The mean value of b for the 16 stars listed by G02 is <b>=0.75±0.19.

In the case of the young stars in clusters, information is available in Figure 40 in G02: not only are the flare energies larger, but the <b> values are also found to be systematically larger at a given stellar luminosity. Thus, among Orion stars, with ages of a few $10^6$ years, <b> ≈ 1.2. Among Pleiades stars (ages ≈ $10^8$ yr), <b> ≈ 1.0. Among Hyades stars (ages ≈ $10^9$ yr), the <b> values become indistinguishable from the values among solar neighborhood flare stars.

In terms of statistics, it must be admitted that the slopes obtained by G02 for solar neighborhood flare stars are based on rather small samples of flares in some cases. The largest sample (for UV Cet) contains 320 flares. Three other stars contain more than 100 flares (YZ CMi, GL 896, and EV Lac). But in two cases (YY Gem, DT Vir), the spectra were constructed based on only 6 flares. And in two other stars (GL 815, EQ Peg), the samples contain less than 10 flares. It is possible that the relatively large value of the r.m.s. deviation (±0.19) in the mean value of <b> = 0.75 are related to the small size of the samples that are available for some stars.

### 2.2 Space-craft data

The sample of flare data has greatly expanded due to the Kepler spacecraft: Davenport (2016) reports almost $10^6$ flares on 4000+ stars. The average number of flares per flare star (≈200) in the Kepler data is larger than the number of flares on any of the solar neighborhood stars used by G02, except for UV Ceti. Davenport has extracted values of a and b (see eq. (1)) for each star: the b values vary widely, from 0.0 to 7.5. At first sight, it seems that the range is too broad to allow us to identify "the" slope of the Kepler flare power-law. However, not all of the slopes are equally reliable: some stars have only a small number (10-20) of flares, and for those stars, b could be subject to large statistical uncertainties.

In order to have more statistical confidence in the b values, we made a first cut through the Kepler sample identifying stars in each of which the number of flares is 200 or more. Using that sample, we find that the mean value of <b> is 0.81. Interestingly, this mean value agrees (within 1σ) with the <b> value reported by G02. However, the r.m.s. deviation in the Kepler <b> value, using the entire database of 4041 stars is large (σ = 0.6), possibly due to the presence of outliers.

A major difference between the Kepler data set and the ground-based flare stars in the G02 sample is that the latter are almost all M dwarfs, whereas Kepler was designed specifically to search for planets around solar-like stars. Could it be that the presence of a mixture of sub-samples of different spectral types is contributing to the large value of σ? In order to test this possibility, we made several cuts through Davenport's table on the basis of mass. In the mass range 0.9-1.1 solar masses (i.e. early G stars), we identified 438 stars: their mean b value was found to be 0.77 with an r.m.s. deviation of ±0.62. These properties overlap with the results obtained from the sample as a whole, suggesting that the overall flare sample is dominated by G stars. Traditional flare stars have masses which are clearly less than the mass of the Sun. When we did a cut through the Davenport data at masses ≤ 0.8 solar masses (i.e. late K and M stars), we found 1358 stars, and these led to <b> = 0.81. However, in this case, the

r.m.s. deviation in b was found to be reduced to ±0.30. To restrict the sample even more towards M dwarfs, we note that among the stars with the best known masses (better than 1% precision), 4 dwarf stars of spectral type M1 are known to have masses of 0.60-0.61 solar masses (Torres et al 2010). Thus suggests that if we make a cut at masses ≤ 0.7 solar masses, we would be including M dwarfs as early as M0 , and eliminate most of the stars with spectral type K or earlier. Making this cut, we identify a sample of 206 stars in the Davenport list. For these M dwarfs, we find <b> = 0.61±0.20. The reduction in σ is noticeable: we suggest that this improvement results from choosing a more homogeneous sample. Moreover, σ is now comparable to the value (±0.19) obtained by G02 in his ground-based sample of 16 flare stars, all of which have spectral type M in the range M1.5-M6.5.

More recently, Kepler data have been independently used to obtain flare energy spectra for a sample of 10 ultracool dwarfs, with spectral types ranging from M6.5 to L0 (Paudel et al 2017). Values of b have been obtained for flares on all 10 stars: the mean <b> for M dwarfs is found to be 0.68±0.21, overlapping well with the <b> value obtained from the sample of M dwarfs reported by Davenport (2016). Within one r.m.s. deviation, the values of <b> obtained for optical flares on M dwarfs by G02 (0.75± 0.19), by Davenport (2016) (0.61± 0.20), and by Paudel et al (2017) (0.68± 0.21), are statistically consistent with one another.

### 3. Energy Spectra in Terms of Flare Frequency f: the Sun

The Sun provides a large sample of data to work with as long as we use X-ray data: the photosphere is dark in X-rays, and so X-ray flares can be detected provided they exceed the noise signals in the satellite-borne detectors. As a historical fact, we note that X-ray data from solar flares actually led to the discovery (in the late 1960's) that the Sun exhibits a power-law energy spectrum before the existence of such a shape was known in stellar flares.

*3.1. X-ray Flares on the Sun*

Drake (1971) reported on a sample of 4028 X-ray solar flares which were detected in the wavelength range 2-12 Å by two separate satellites over a 2.2-year interval (July 1966-September 1968).  Drake derived a differential spectrum with α = 1.44 ±0.01. In terms of eq. (1), Drake's results correspond to b = 0.44±0.01. Drake's value of b for solar X-ray flares does not overlap (within 1σ) with the value of <b> obtained by G02 among a sample of optical flares. However, Drake's value is within 1σ of the slope for M dwarf flares in Kepler data.

Kasinsky and Sotnicova (2003: hereafter KS03) used GOES satellite archives to obtain X-ray data (1-8 Å) extending over three solar cycles, from 1972 to 2001. KS03 analyzed the data in terms of an integral energy spectrum as in eq. (1) above**.** (A similar, but somewhat smaller, GOES data set, extending from 1976 to 2000, was also used by Veronig et al 2002, and analyzed in terms of a differential energy spectrum.) KS03 reported 153 flares per year at their quietest solar minimum (1976), and 4005 flares per year at their most active solar maximum (1981). That is, the GOES annual flare rate varies by a factor of up to 26 during the solar cycle. The energy spectrum for a particular year (plotted in Fig. 1 of KS03), showed that all three regimes (i)-(iii) are present in the data: when KS03 derive a value of the slope b, they use data in regime (iii). Individual values of b were obtained for each year: the values of b ranged from as small as 0.50 (in 1974, near solar minimum) to as much as 0.80 (in 2000, near solar maximum).

In the global sample, containing 56,600 flares, KS03 found the mean value of <b>(1972-2001) to be 0.666±0.005: the precision of the global <b> value is remarkable. Converting from differential to integral spectra, we find that the results of Veronig et al (2002) correspond to <b> = 0.88±0.11: this is statistically steeper than KS03 found, although it overlaps with the slope of 0.80 in 2000 (near maximum). It is possible that the differences between the analyses of KS03 and Veronig et al may be due to the fact that Veronig et al did not include the years 1972-1975, when the KS03 slopes were among the shallowest.

In view of the larger size of the KS03 dataset, as well as the more extensive time-period to which it refers, and in view of the smaller error bar of the KS03 result, we shall assume in what follows that the flare energy spectrum in the Sun in regime (iii) is best described by a power-law of the form in eq. (1) with a value of = 0.666±0.005 for <b>. In a statistical sense, the KS03 value of <b> overlaps (within 1σ) with the values of <b> obtained by G02 and by Davenport (2016) in their studies of M dwarf flare stars.

This value of <b> obtained by KS03 using GOES data is, in a statistical sense, significantly different from the value obtained by Drake (1971). Veronig et al (2002) also commented on this difference between their spectral slope (based on GOES) and Drake's. It is possible that these difference are due to the very different orbits for the spacecraft: Drake used satellites in highly eccentric orbits reaching as far as the Moon and beyond. In contrast, KS03 and Veronig et al (2002) used data from satellites in circular geostationary orbits in the equatorial plane at a fixed radial distance (≈42,000 km) from the Earth's center. The ambient populations of energetic particles were quite different for Drake's satellites and for GOES.

### 3.2. Correlations between Energies of X-ray Flares and Optical Flares?

Is there any reason to expect correlations between X-ray flares and optical flares? The answer depends on whether or not there is a common physical process which contributes to both types of flares. A flare is triggered when magnetic energy is released in a coronal loop, generating hot plasma which emits X-rays. In order to emit optical radiation, a process is required to transport some of the released energy down to the photosphere (or close to it) so that optical photons ("white light flares") can be generated. Electron beams or thermal conduction can do the transport, channeled by the field lines down to the foot-points of the loop. In the case of thermal conduction, the ratio of X-ray to optical emission is proportional to the ratio of the time-scales for two cooling processes: radiation and conduction (Mullan 1976), i.e. the X-ray flux is proportional to the optical flux. Kuhar et al (2016) have demonstrated that all solar flares of GOES class M5 or larger exhibit white light emission with a flux that is correlated with hard X-ray flux: smaller flares may also have white light emission, but the radiant flux from the photosphere is so bright that the flare light cannot be detected reliably.

### 4. Energy Spectra in Terms of the <u>Time Interval</u> T between Flares

The time interval T between flares is of order 1/f. As a result, in a regime where power-law behavior appears in a frequency plot, we can say that a power-law relationship also exists between E and T:

Log T = -a + b*log(E)     (2)

Thus, if we write that the energy E is related to T by $E \sim T^s$, then the index s is related to the quantity b in eq. (1) by the relationship s = 1/b. Inserting the KS03 value of <b> for X-ray flares on the Sun, this leads to

$E \sim T^s$, with s = 1.50±0.01 (solar X-rays)    (3)

This means that if the Sun can "hold off" for a time interval T without flaring, then at the end of the interval T, an X-ray flare with total energy of E or greater will occur. And the value of E scales as $T^{1.5}$.

How might we understand the scaling $E \sim T^{1.5}$? Presumably, in the course of the interval T, the Sun is building up energy which will be released in the form of X-rays when a flare eventually occurs. In physical terms, we ask: is there a process at work in the Sun which, in regime (iii), builds up energy over time T in such a way that if the system eventually flares, the released energy scales as $T^{1.5}$? We have already pointed out (Section 1) that a power law with slope -1.5 fits most of the solar flare data in Fig. 1.

What about regime (i)? At the largest energies, in the cases of the Sun, AD Leo, EQ Peg, and UV Cet, the flare spectra shown in Fig. 1, as well as the KS03 data (their Fig. 1), show a flattening. That is, as the interval T between flares reaches its largest values, E does not increase as steeply as $T^{1.5}$. Instead, if we fit a power-law dependence, the largest flares will exhibit $E \sim T^s$ where the numerical value of s is smaller than 1.5. Inspection of Fig. 1 shows that, for the stars with the largest flare energies (young stars in Orion) the flare energy spectra have slopes with values of s which are ≈1, i.e. definitely shallower than the slopes in regime (iii).

If we can identify a physical process which leads to a power-law energy spectrum, it may improve our understanding of flares. G02 has summarized several models which were previously proposed to do this, but each of those models was subject to some limitations (as described by G02). In the present paper, we offer a new proposal to explain the existence of power laws with preferred values of the slope.

## 5. Flare Triggering by Instability in a Magnetic Loop

In this paper, we examine the flare energy spectrum in the context of flares which occur in a coronal loop. The Skylab data obtained in the 1970's showed that there are two major classes of flares (e.g. Priest and Milne 1980): (i) compact flares, which are confined to a single coronal loop; (ii) 2-ribbon flares which involve typically interactions between two or more loops. In general, the 2-ribbon flares involve larger releases of energy, while the compact flares are associated with smaller energy releases. In terms of the standard flare energy spectrum, and for purposes of the present paper, it seems plausible to assume that the power law portion of the spectrum, in the lower energy range, refers to compact flares (i.e. the smaller class of flare). The flares in which we are interested here are those which occur in a single loop.

A sketch of a coronal loop is shown in Fig. 2 (adapted from Kittinaradorn et al 2009 [hereafter KRM]: used with permission). The colored parallelograms were used by KRM for their discussion of current sheets in the loop, but they are not relevant in the present paper. For present purposes, the relevant aspects in Fig. 2 are: (i) each foot-point intersects the lower solar atmosphere in a region which is roughly circular with diameter L, and (ii) inside each foot-point, there exist multiple smaller structures, each with a diameter of roughly λ. To help distinguish these smaller structures, KRM illustrated them as white against a red background, but the colors have no physical significance. However, KRM were interested in explaining the phenomenon of solar "moss", which has a patchy appearance in ultraviolet emission: as a result, their illustration indicates that the smaller structures do not occupy all the available area inside the foot-point. KRM attributed the structures in each foot-point to patchy heat *conduction from above*. In contrast, in the present paper, we shall consider small structures inside each foot-point as having a very different origin: we consider them as arising from heat *convection from below.* In this case, the small structures we consider will occupy *all* of the available area inside each **foot-point, leaving no gaps. We shall quantify the length scales L and λ in Sections 7 and 8 below.**

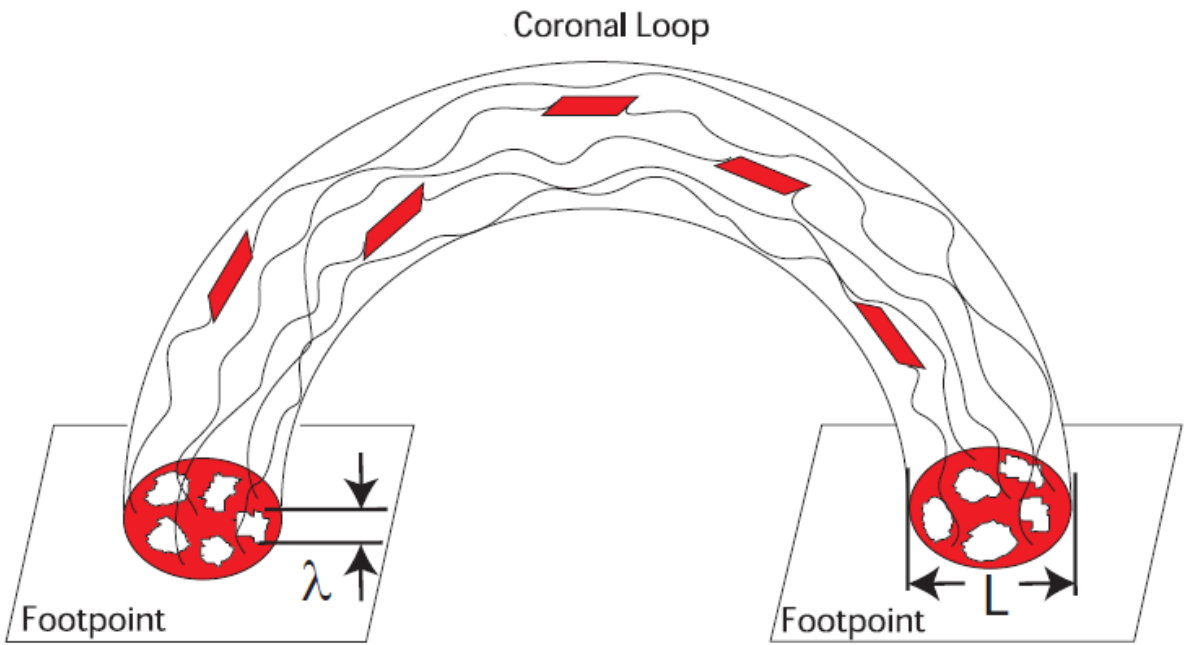

Figure 2. Perspective sketch of a coronal loop and its intersection with the lower solar atmosphere: each foot-point of the loop is illustrated as a circle located on a rectangular horizontal cut through the photosphere. (Adapted from KRM, with permission.) The length L represents the diameter of a foot-point at the level where it intersects the photosphere. The length λ represents the diameter of one of the photospheric granules which happen to lie inside the foot-point.

A large number of physical mechanisms have been proposed to explain flares in terms of magnetic energy release (e.g. for an overview, see Spicer and Brown 1981). Here we focus on flares which occur in a coronal magnetic loop which has been driven to an unstable condition. In this context, we consider two models, one due to Alfven and Carlqvist (1967: hereafter AC), and the second due to Hood and Priest (1979: hereafter HP).

In the AC model, a current J flows in a current channel in the solar atmosphere. (The channel is in some cases a coronal loop.) The value of J may reach a critical stage: the current $J = ne(v_e - v_i)$ may become so large that in order to carry it, the local electrons have to move at speeds relative to the ions which exceed a critical value (related to the thermal speed of the electrons). In these conditions, an electrostatic instability sets in and the local resistance R suddenly increases to large values. The large R value leads to rapid dissipation of the current at a rate proportional to $J^2R$, and this manifests itself as a flare. In the context of a coronal loop, the value of J is related to the curl of the field: $J = (c/4\pi)$ curl(H). Therefore, one way to increase the current, driving it to larger and large values, is to twist the loop more and more as time goes by. When the twist reaches a large enough value, the associated current goes critical. In this sense, if a loop can be twisted enough, a flare will occur.

In the HP model, an MHD analysis of the stability of a twisted coronal loop leads to the conclusion that such a loop can be (initially) stable over a broad range of conditions. But if the amount of twist along the loop exceeds a critical value, the loop becomes subject to a kink instability: reconnection can occur at the kink, releasing magnetic energy, and causing a flare. The critical twist is found to be between $2\pi$ and $6\pi$, depending on local conditions.

6. **Driving a Coronal Loop to Become Unstable**

In the context of flare activity, the most remarkable empirical characteristic of solar coronal loops is their surprisingly long-lived MHD stability. That is, for the most part, solar loops surprise us because they do **_not_** flare. Many loops which have been monitored by X-rays detectors on (e.g.) Skylab or Yohkoh are observed to remain stable for days on end.

The striking aspect of this empirical result is that the MHD time-scale for any given loop (i.e., the time required for MHD fast-mode waves to cross an individual loop) is measured in seconds (Van Hoven 1981). The existence of such short time scales suggests that, in principle, the loop should survive for only a matter of seconds. And yet solar loops survive for days. What causes the loops to possess such a great degree of stability? The answer is: the foot-points of all coronal loops are rooted in the dense gas of the photosphere. Because of the high density and the relatively high electrical conductivity of photospheric gas, the magnetic field lines in a flux rope which arches upwards (as in Figure 2) high enough to reach the corona remain nevertheless "tied" to the photosphere because the field is "frozen in" to the gas. It is this "line-tying" that keeps the coronal loop stable under a broad range of physical conditions.

In order for a flare to occur, ambient processes must be such as to overcome the effects of line-tying sufficiently that the loop is eventually driven unstable. In both the AC and the HP scenarios of a flare, the key factor is the occurrence of a *twisting* process which, when applied to a coronal loop, can eventually lead to instability. How might a twisting effect be generated in a coronal loop on the Sun or in an M dwarf? To answer this quantitatively, we first outline some of the parameters of a loop.

## 7. Relevant Loop Parameters

We define a coronal loop in terms which are guided by the detailed images of the Sun which were first obtained in X-rays by Skylab (e.g. Eddy 1979). A coronal loop is a magnetic flux tube which emerges from the sub-surface layers at a particular region in the photosphere (the "foot-point"), rises up to a finite altitude h above the surface, and then bends over to re-enter the solar surface at a second foot-point. The total distance along the loop from one foot-point to the other is S. Each foot-point has a certain linear diameter L on the solar surface (see Fig. 2).

In the case of a loop which is stable enough to live a "long" time (where "long" means long enough to be a potential site for a flare), the value of h can probably not depart significantly from the value L: if a loop were to be such that it reaches the limit h>>L, then the loop would be so elongated (in the vertical direction) that there would be a danger that the loop might pinch off at some point and disconnect from the Sun. In the opposite limit, i.e. where the loop has h<<L, the loop would emerge from one foot-point in a vertical direction, but then would very quickly bend over towards the horizontal. The loop would have to remain "squashed" close to the stellar/solar surface (along most of its horizontal length) until it approached the other foot-point: at that location, another sharp bend would be necessary in order for the loop to re-enter the photosphere in a vertical direction. In such a loop, strong currents would build up at the bends (where curl(H) takes on very large values) and should be quickly dissipated.

This line of reasoning suggests that h is not likely to differ greatly from L. If the loop defines a torus, then the major radius of the loop is h, such that the loop-length $S = \pi h$. Assuming that each foot-point can be approximated as a circle (as in Fig. 2), the minor radius of the torus would be $r = L/2$. The aspect ratio A of the torus, defined to be the ratio of major to minor radius, has the value $A = 2h/L$. In solar coronal loops, empirical values of A typically range from a few to perhaps 10. In terms of loop length S, the values of L are expected to be of order $2S/A\pi$: with $A \approx 3$, we expect L to be of order S/5. Visual inspection of solar images in X-rays suggest that coronal loops which undergo detectable flaring have lengths of typically S = a few times $10^9$ to a few times $10^{10}$ cm. These values suggest that typical values of L in the Sun are of order $10^{9-10}$ cm (e.g. Mullan, 2009). We shall use these L values to interpret the empirical spectra of X-ray flares in the Sun.

## 8. Loops and Granules in the Sun

In the present context, a relevant parameter is the ratio of L to the size of the granules which are observed to exist at all parts of the solar surface. The granules are manifestations of convective flows in a quasi-cellular pattern (see, e.g. Mullan 2009). Granules consist of (i) a central bright region where hot gas rises from below, (ii) dark lanes around the outer edge of the granule where cooler gas sinks beneath the surface, and (iii) gas flowing horizontally from center to edge. Although granules have a range of sizes, the distribution of granule diameters has a peak at a length-scale λ which is found to be of order $10^8$ cm (Mullan 2009). Each granule survives for a period of time τ which allows its material to circulate (roughly) once around the convection cell. Thus, τ is of order λ/v, where v is the mean speed of flow within the cell. With v observed to be in the range 1-3 km/sec, the value of λ/v is expected to be of order 300-1000 sec. At the solar surface, observations do indeed indicate that individual granules survive for periods of order 5-10 minutes, consistent with the λ/v estimates.

Using the above estimates of L and λ, we can obtain a quantity which plays an essential role in the present paper, namely, the ratio of L to λ. In the Sun, the empirical information indicates that L/λ is of order 10-100. We will need this ratio in eq. (5) below.

The foot-points of a solar coronal loop are therefore immersed in a medium where many small convective elements are constantly performing their convective flows, both inside and outside the foot-point. Inside the foot-point, individual magnetic flux ropes belonging to the loop will intersect the photosphere. Some of the flux ropes have diameters which are small enough to be contained entirely within a single granule. Because of high electrical conductivity in the gas, such a rope will be subject to continual pushing or dragging ("jostling") by the horizontal flows inside individual granules. Flux ropes which are large enough to have foot-points covering more than one granule will be subject to reduced amplitudes of jostling. Since the field lines of a flux rope intersect the photosphere essentially vertically, the field is effectively unaffected by the vertical motions of gas in the bright center or in the outer dark lanes. We consider that it is the horizontal flows associated with granular motions which have the potential to twist a loop in preparation for a flare.

As an illustration of the process which is at work, we show schematically in Figure 3 an area of the solar photosphere where the circular foot-point of a coronal loop happens to lie. The foot-point has a diameter of L (the same quantity as is illustrated in Fig. 2). Inside the foot-point, and surrounding it, there are granules, each with diameter λ: examples of two of these granules inside the foot-point are illustrated in Fig. 3. (There are actually many more than these 2 inside the foot-point, but for clarity we focus here on just 2.) At any instant of time, a particular granule will drag the field line which happens to be passing through that granule. We will focus on two field lines, with foot-points labelled S1 and S2 in Fig. 3, in what follows. The zig-zagging lines attached to S1 and S2 are meant to suggest that the nearest granules (i.e. the two which are illustrated) drag each field line in a series of horizontal steps as time goes by. The mean length of each step is of order λ.

It is important to note that the horizontal flows which give rise to the zig-zag motion of a particular field line are by no means constant in time. On the contrary, since each granule survives for only a finite time, any particular field line will be dragged in a certain direction as long as a particular granule survives, i.e. for a time of order 5-10 minutes, and then when a new granule forms around that field line, the line will (in all likelihood) be dragged in a different direction. As long as each granule survives, the field line will be dragged across a distance of order λ. As a result, the magnetic field lines in the foot-point of a loop are subject to a random walk due to the ever-present and ever-changing direction of the horizontal flows which "buffet" any particular field line.

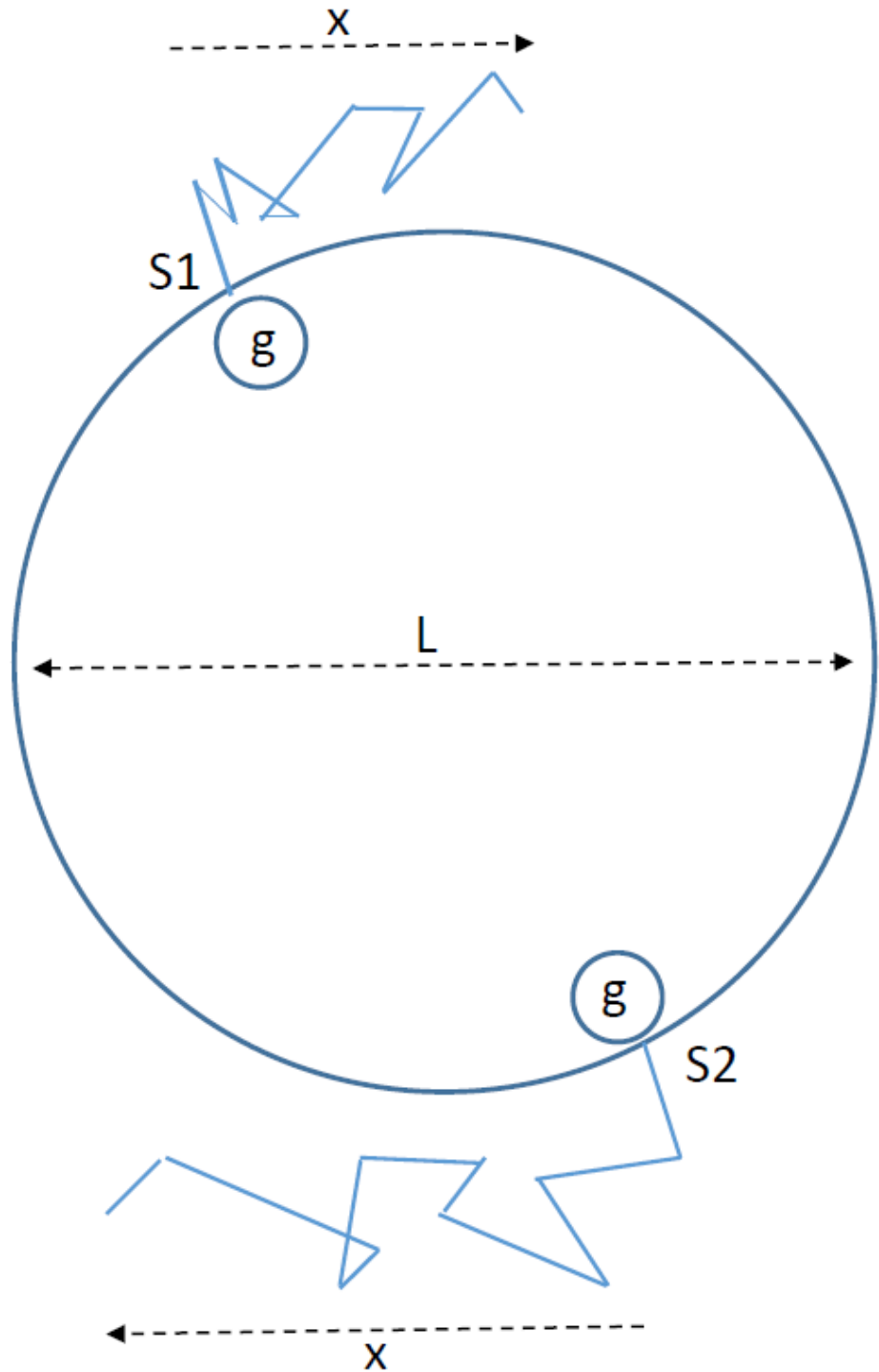

Figure 3. Sketch to illustrate the basic process considered in this paper: how can the foot-point of a loop be twisted by granulation? The circle with diameter L represents a horizontal cut through the loop at the altitude where the loop passes through the photosphere. A field line at the point labelled S1 on the foot-point is subjected to granules which, over time, happen to move the field line in a series of steps mainly

towards the *right* by a distance of order x from its starting position. Each step corresponds to the dragging effects which are imposed on the field line during the lifetime of a single granule: each step has a length of (on average) λ, the diameter of a granule. At the point S2, roughly diametrically opposite S1 on the foot-point, another field line is subjected to dragging from other granules: these granules, over time, happen to move the field line in a series of steps (each of length λ on average) mainly towards the *left* by a distance of order x from its starting position. We suggest that in situations where 2x becomes as large as L, the loop in effect undergoes enough twisting to drive the loop unstable.

A detailed physical modeling of the MHD processes which lead to a flare in a loop that is rooted in granules requires extensive numerical work. A recent example of such work is provided by MHD simulations of a coronal volume (Knizhnik et al 2017) where transient, cellular boundary flows are designed to model the processes by which the photosphere drives the corona in such a way that magnetic helicity is conserved. In the present paper, we adopt a much more simplified algebraic approach to understand how granular driving may lead to a twist in a single loop.

### 9. How might a Coronal Loop Acquire a Critical Twist?

How much twisting is required to drive a loop unstable? According to HP, the twist along the loop must exceed a critical value of order $2\pi$ or more. The ideal way to achieve a twist of $2\pi$ would be to have one foot-point rotate coherently through $2\pi$ without a compensating rotation at the other foot-point. But it is hard to imagine that one foot-point of a loop, immersed in the turbulent convection, would be able to rotate as a coherent unit. Instead, a group of field lines which intersect a particular granule will undergo a random walk with individual step size of order λ. Field lines which intersect a different granule will be dragged in a different direction, but will also experience a random walk with step size of order λ.

Of course it is unrealistic to expect that the Sun will arrange for a coherent twist of the foot-point. Instead, the most we can hope for from a random walk is that from time to time, the coming and going of granules which jostle against one location (S1) of the perimeter of the foot-point will lead to the following effect: after the random walk has been going on for a time period $t_r$, i.e. after the occurrence of N steps of the random walk (in which each step is of length λ), magnetic field lines which originally were located at S1 will find themselves displaced in some direction by a mean distance x from their starting position. (See Figure 3.) The theory of 1-dimensional random walk indicates that the overall displacement x after N steps is related to the length λ of each step by the formula $x \approx \lambda\sqrt{N}$ (e.g. Wolfram 2000). Therefore, if we wish to achieve a systematic displacement of amplitude x in a random walk, we need to allow enough time to elapse to allow N steps to occur, where

$N \approx O(x/\lambda)^2$      (4)

While this systematic shift is occurring in S1, there are also jostlings going on all around the foot-point of the loop, leading to other random systematic shifts in other segments of the foot-point. Referring to Fig. 3, consider the segment S2 which is initially diametrically opposite S1 on the perimeter of the loop foot-point. If it were to happen (although this case is *not* illustrated in Fig. 3) that S2 is also displaced by an amount x during $t_r$ in the *same* direction as the shift of S1 **,** then the foot-point would simply undergo a linear shift as a more or less monolithic structure. Such a loop would exhibit no tendency to become twisted: instead, it would simply undergo lateral displacement.

But suppose the systematic shift of S2 during the time interval $t_r$ is *not* in the same direction as the shift of S1. Suppose, instead, that the shift of S2 happens to be in the *opposite* direction to the shift of S1. (This is the situation that is sketched roughly in Fig. 3.) Then the perimeter of the foot-point will be stretched by a distance +x on one side (S1) and simultaneously by a distance –x on the other side (S2). Such a combination, with its net shift of 2x between the initial location of S1-S2, and the final location of S1-S2, would have the effect of causing a non-zero twist in the loop. The geometry of the situation (see Figure 3) is such that the angle of the twist is estimated to be of order 2x/L radians. To be sure, having S1 and S2 undergo exactly opposite systematic displacements (as shown in Fig. 3) in time $t_r$ is unlikely to happen often in the turbulent environment of granulation. But as long as the shift of $S_2$ is not in the same direction as the shift of $S_1$ , the outcome will be that, after a finite time $t_r$ , the loop will undergo a finite twist, rather than simply a lateral displacement.

We are most interested in what happens when, at some location, the displacement x grows to a value of order L during a **finite** time interval $t_r$. In such a case, the twist (of order 2x/L) could grow to a value of order 2 radians. This is approaching, within a factor of a few, the limiting twist which will drive a loop unstable (see HP). The essential ansatz in the present paper is this: the process of twisting caused by random jostling of the perimeter of the foot-point can (if 2x → L) drive the loop to instability, i.e. to flare. Quantitatively, the mean time interval between flares is determined, in our model, by the time that must elapse before x has increased to a value of order L. We now turn to an estimate of the time required to do this.

## 10. Time-scale T Required to Drive a Solar Loop Unstable

How much time will it take for a critical twist to be built up in a loop in the solar corona? The random walk we have been considering consists of "steps" , each of which takes a finite time: this finite time is associated with the lifetime of an individual granule τ = 5-10 minutes. Inserting x=L, the number of steps N required to accumulate a shift of L will be of order N = $(L/\lambda)^2$. Inserting the above estimates of L/λ, this leads to N = $10^2 – 10^4$ . With each step lasting a time τ, the time interval between flares on a loop is expected to be of order

$T = (L/\lambda)^2 \tau$ (5).

Inserting appropriate values, we predict that T should lie within a range from roughly 500 minutes to $10^5$ minutes, i.e. between several hours and as long as tens of days. On a frequency scale where the units are (hours)$^{-1}$, these predictions suggest that the maximum frequency f of solar flares should occur at log f ≈ -1, while the minimum frequency should occur at log f = -3.2. Inspection of Fig. 1 indicates that such a range of frequencies overlaps well with the empirical boundaries of the energy spectra obtained for flares in the Sun. In view of this, we suggest that granule-driven twisting of a coronal loop provides a plausible mechanism for understanding in general terms the frequency of solar X-ray flares.

In this paper, we would like to stress that our interest is specifically in the *small-scale* motions associated with granules as the source of driving a loop towards instability. However, in the Sun, another process also leads to magnetic field distortion on a large scale: differential rotation (DR). (This process plays a fundamental role in the Babcock (1961) theory of the solar dynamo.) In principle, loops in the Sun which inter-connect active regions might have foot-points that are far enough apart to be affected by DR: such loops have lengths of up to 25 heliographic degrees, although some cross-equatorial loops can be as long as 37 degrees (Bray et al. 1991). Transient brightenings have been reported in some interconnecting loops, suggesting that they may be subject to some kind of instability, but evidence suggests that flares are not at work in most of these brightenings (Svestka and Howard 1979). The loops in which we are primarily interested in the present paper are so short (typically a few times $10^9$ cm, i.e. a few heliographic degrees) that even if they happened to be aligned strictly north-south, the DR between one end of the loop and the other would be so small that driving the loop to instability would be unlikely. Only if we were to consider loops extending over many times $10^{10}$ cm in length would be need to take into account the effects of DR.

Furthermore, an anonymous referee has pointed out that in contrast to G dwarfs (where differential rotation is significant), M dwarfs have rotations which are more solid-body like. For example, differential rotations of 0.0063 rad day$^{-1}$ and 0.012 rad day$^{-1}$ have been reported for M dwarfs (Davenport et al 2015): these are up to an order of magnitude smaller than in the Sun (0.055 rad day$^{-1}$). And yet the specific flare rates in M dwarfs are observed to be characteristically higher than in G dwarfs (e.g. G02). This suggests that differential rotation probably does not play a dominant role in driving flares in M dwarfs.

## 11. Scaling Between Energy E and the Time Interval T between Flares

Following G02, we use the results of Pustil'nik (1986), who ascribes the energy of a flare to the conversion of some of the available magnetic energy inside a certain volume V into a form which can generate radiation. In a region where the local field strength is H Gauss, the initial magnetic energy density is equal to $H^2/8\pi$ ergs cm$^{-3}$. A reduction in this energy (due perhaps to magnetic reconnection) is what is presumed to power a flare. Therefore, the energy released in a flare is expected to be

$$E(f) \approx (\Delta H^2 /8\pi)V \qquad (6)$$

where V is the volume inside which the field is reduced, and $\Delta H^2$ is the reduction in $H^2$ which occurs in order to generate the flare. In a loop with foot-point diameter L, and length S which is related to L by S~L, say S = 5L (see above), the geometric volume of the loop is roughly V = $(\pi/4)L^2$ S ≈ $(5\pi/4)L^3$. However, the geometric volume of the loop may not be the same as the volume inside which the magnetic field undergoes a process of flare-related reduction in strength: the reason is that the field strength H cannot necessarily be treated as a constant throughout the loop. In any solar structure, the value of H decreases with increasing height above the photosphere: there exists a characteristic height z (the "magnetic scale height") over which H decreases with increasing height. If we happened to be dealing with a global dipole (poloidal) field, the H value would decrease as $1/r^3$, where r is the radial distance from the center of the star. In such a case, the characteristic value of z would be comparable to the radius of the star. But in flares, we do not deal with the global dipole: instead, we are dealing with the field in an active region where locally strong toroidal fields originate in localized flux tubes which rise up from a dynamo source below the surface. In an active region, the strong local fields can be approximated in terms of a "buried dipole" (e.g. Mullan 1979) located at a finite depth $h_o$ below the surface: in such cases, the H value falls of as $1/d^3$ where d is the distance from the buried dipole. If the latter is close to the surface, the decrease in H with height is much more rapid than $1/r^3$ : in such cases, the magnetic scale height z is much smaller than r.

The total flare energy depends on how the vertical extent of the loop compares to z. Since the maximum altitude of a loop is proportional to L (see Section 7), two limiting cases are relevant.

In the event where L<<z, we are dealing with a small loop: in this limit, we can treat the H value as more or less constant throughout the loop. In such a case, the flare energy can be generated throughout the entirety of the geometric volume of the loop. In that case, we can consider the flare to be effectively 3-dimensional, and the flare energy can be written as

$E(3) = (\Delta H^2 /8\pi) (5\pi/4)L^3$ (7)

The quantity E(3) refers to the energy which is released when a flare occurs on smaller loops, i.e. in the limit of smaller flares. For convenience in what follows, such (smaller) flares will be referred to as 3-D structures.

In the opposite limit, where the loop size L is large compared to z, the flare energy is effectively generated only within the volume where the magnetic field is strong. At larger heights, where the H value decreases, there is less magnetic energy available per unit volume to contribute significantly to the flare energy. In effect, the volume which contributes to flare energy becomes $V(2) = (\pi/4)L^2z$, where the effective volume contributing to the flare is more or less 2-dimensional. (For an empirical discussion of this 2-D limit in the case of large solar flares, see Aschwanden [2016, his Fig. 14(c)], where the initial volume is found to scale with L to a power which is *not* equal to 3, but instead has a value of 1.98±0.02. In Aschwanden's work, our parameter z is replaced by the local scale height of gas density in the Sun's atmosphere.) This leads to a flare energy given by

$E(2) = (\Delta H^2 /8\pi) (\pi/4)zL^2$ (8)

The quantity E(2) refers to the energy which is released when a flare occurs on a larger loop, i.e. in the limit of larger flares. The paper by Aschwanden (2016) deals with the 399 largest flares detected by Solar Dynamics Observatory in the first 3.5 years of the mission: the flares in the sample were classified as M-class and X-class. Therefore, Aschwanden's sample refers to flares in which the loop size L is expected to be large compared to the characteristic scale z: these flares have energies which are described by eq. (8).)

Using eqs. (7) and (8), we can now determine how we expect the flare energy to scale with the time interval T between flares. Given that $T = (L/\lambda)^2 \tau$ (see eq. (5)), we can write

$L \approx \lambda T^{0.5} /\tau^{0.5}$ (9)

Inserting eq. (9) into eq. (7), we find that for 3-D flares, the energy depends on T according to

$E(3) = (\Delta H^2 /8\pi) (5\pi/4) \lambda^3 \tau^{-1.5} T^{1.5} = 0.2 \lambda^3 \tau^{-1.5} \Delta H^2 T^{1.5}$ (10)

This indicates that, if λ, τ, and $\Delta H^2$ do not vary systematically from one flare to the next in any particular star, then we can draw the following conclusion for that star: 3-D flares (on smaller loops) are expected to exhibit E values which scale as $T^s$ where s = 1.5.

We consider it noteworthy that the power-law index obtained in eq. (10) (i.e. 1.5) agrees well with the empirical index of 1.50±0.01 reported by KS03 for smaller solar flares, i.e. those in regime (iii) (see Section 4 above). We have already illustrated (Section 1) that the solar flare spectrum in Figure 1 is fitted over most of its range by a power law with index 1.5: thus, in the context of our model, we suggest that most of the solar flares plotted in Fig. 1 can be assigned to our 3-D category.

For 2-D (larger) flares, for which the energies are given by the expression for E(2), we insert eq. (9) into eq. (8) and find

$$E(2) = (\Delta H^2 /8\pi) (\pi/4) z \lambda^2 \tau^{-1} T \qquad (11)$$

Thus, if λ, τ, z, and $\Delta H^2$ do not vary systematically from one flare to the next in any particular star, then we can draw the following conclusion: 2-D flares (on larger loops) are expected to exhibit E values which scale as $T^s$ where s = 1.0. The index we find for 2-D flares is definitely shallower than the index which is relevant for the (smaller) 3-D flares in regime (iii). We suggest that the largest flares in Fig. 1, i.e. those in regime (i) where the power-law slopes become shallower, can be assigned to our 2-D category. Quantitatively, we have already illustrated (Section 1) that the spectrum for the star with largest flare energies (labelled Ori1 in Fig. 1) is fitted well with a power law with index 1.0, as we derive for 2-D flares. But even in the case of solar flares, the flattening which occurs in Fig. 1 at the highest energies, is accommodated better with a power-law index of 1.0 than 1.5. Similar remarks can be made about the two other stars (UV Cet, AD Leo) in Fig. 1 which show flattening at the highest energies. This leads us to suggest that the flares on Ori1, as well as the largest flares on the Sun, on UV Cet and on AD Leo may be assigned to our 2-D category.

In Section 15 below, we will examine the plausibility of the assumption that λ, τ, z, and $\Delta H^2$ do not vary systematically from one flare to the next in any particular star.

### 12. Magnitudes of Flare Energies in the Sun

As well as checking the power-law index of the flare energy spectrum, we now turn to evaluating the magnitude of the energy which is expected to be released in solar flares in compact loops. Let us consider flares which occur at intervals of T = 100 hours = 3.6 x $10^5$ seconds. According to G02 (Fig. 38), such flares in the Sun have energies of order $10^{29}$ ergs. Let us use eq. (10) to see if solar parameters are consistent with these quantities. Inserting λ = $10^8$ cm and τ = 300-600 seconds, we find consistency provided that $\Delta H^2$ has a numerical value of order 20-50 in units of (Gauss)$^2$, i.e. H ΔH ≈ 10-25 $G^2$. Empirical studies of magnetic field strengths in coronal active regions which emit radio and X-ray radiations (Schmelz et al 1994) indicate that the coronal field strengths H range from as small as 30 G to several hundred G. Therefore, reductions in field strength ΔH of no more than 1 G (or less) would suffice to provide the magnetic energy which is observed in solar flares which occur once every 100 hours. That is, there is no need to demand that a significant fraction of the available magnetic field be destroyed in order to power a solar flare: even a weak amount of reconnection suffices to power the flares.

Overall, we see that the concept of compact (i.e. small, 3-D) solar flares originating in the twisting of coronal loops until the loops become unstable replicates two empirical quantities: (i) the observed power-law index of the flare spectrum, and (ii) the magnitude of the flare energies in terms of plausible magnetic fields.

How can we understand the observed variations in flare rate across the solar cycle? The number of flares which are detected in any given year is observed to vary by a factor of up to 26 between solar minimum to maximum (KS03). We can use eq. (10) to understand this in terms of cyclic variation in solar fields. At solar maximum (minimum), the field strengths H in flare sites are expected to be strongest (weakest). Therefore, other things being equal, the flare energies E(3) are expected to be largest (smallest) at solar maximum (minimum). Because a detection threshold always exists in any detector because of background counts (i.e. "dark current"), the detection of large flares is always easier than when the flares are small. Moreover, at solar minimum, a larger supply of galactic cosmic rays reach the Earth (e.g. Mullan 2009, p. 337), thereby contributing to enhanced background in space-borne detectors. We suggest that the lower detection rates of X-ray flares at solar minimum might be attributable to a combination of smaller E(3) values and simultaneously higher noise levels in the detectors. This combination has the effect that more flares are expected to find themselves in regime (ii) (see Section 1), where they are "lost in the noise". At the same time, the flares which do have enough energy to be detected in these noisy conditions inevitably have larger energy: as we have seen (Section 1), flares with larger energies do not occur as frequently as low-energy flares. Therefore, the flares which are actually detected at solar minimum will occur at a slower rate: we suggest that this can account for the smaller flare rate reported during solar minimum. However, the flares which do have enough energy to rise above the noise (and therefore be detected as regime (iii) events) will still obey eq. (10): the power-law index which describes E as a function of T, will still be 1.5, even though the overall number of detected flares is smaller.

### 13. Magnitudes of Flare Energies in Solar-Neighborhood Stars

Up to this point, we have used parameters for magnetic loops in the Sun. Now we turn to flare stars, where various parameters may take on values that are different from those in the Sun.

Empirically, the results presented by G02 in his Fig. 38 demonstrate that, at a given frequency, flares in solar-neighborhood stars exhibit energies which exceed the solar flare energies at the same frequency (e.g. once every 10 hours) by factors of 2-5 orders of magnitude. In view of eq. (7), it is possible to retain the unstable coronal loop to account for such flares provided that the product $L^3 \Delta H^2$ is larger in a flare star than in the Sun by factors of $10^{2-5}$.

Is there any evidence to support such large increases in M dwarfs? As regards lengths of loops, Mullan et al (2006) reported that the lengths of flaring loops increase to values as large as twice the stellar radius (i.e. several times $10^{10}$ cm) among stars later than M2-3. These loops are significantly longer than the loops we have discussed for the Sun, where the largest value of L was estimated to be of order $10^{10}$ cm. As a result, the $L^3$ values in flare stars may be up to $10^{2-3}$ times the solar value. Moreover, it is not merely the lengths of loops which are larger in flare stars: field strengths in flare stars have also been found to be significantly stronger than in the Sun. E.g. in UV Cet, WX Uma, and Wolf 47, dipole fields with magnetic flux densities of 6-8 kiloGauss have been reported (Shulyak et al 2017). Such fields are at least 100 times stronger than the dipole fields which occur at the poles of the Sun. Therefore, even if ΔH in flare stars is as small as in solar flares, the flare energy (proportional to the product HΔH L$^3$) can be of order 10$^5$ times larger than in the Sun.

We conclude that as far as flare energies are concerned, flares in regime (iii) in solar neighborhood stars can be considered as not inconsistent with the hypothesis that flares on such stars are quantitatively consistent with compact loop being driven to twisting instability.

As regards the time interval T between flares, eq. (5) above indicates that the larger values of L in flare stars will contribute to increasing T values in stellar flares. But the parameters λ and τ are also expected to change from their solar values. If granules sizes are determined by the local pressure scale height H~T/μg, then as we go down the main sequence, decreasing T and increasing g and μ will cause λ to decrease by factors of a few as we go from the Sun (T=6000 K) down to mid-M dwarfs (T=3000 K). Therefore (L/λ)$^2$ will increase by factors of perhaps 10-100 as we go from the Sun down to mid-M dwarfs. On the other hand, Mullan (1984) has argued that the values of τ should decrease from the solar value (of several hundred seconds) to about 100 seconds at T(eff) = 4000 K, and decrease further to 40-50 seconds at T(eff) = 3000 K. According to Rajpurohit et al (2013), these T(eff) values correspond to spectral types of M0 and M5, a range which covers most of the stars for which G02 lists a and b values. Combining the 10-100 times increase in (L/λ)$^2$ with a decrease by a factor of order 10 in τ, eq. (5) suggests that time intervals between flares in M dwarfs should be comparable to, or somewhat larger than, those in the Sun. In the plot of G02 (Fig. 38), there is certainly overlap between the time intervals between flares on M dwarfs and those in the Sun. It is not obvious that there is any systematic increase in τ among flare stars, unless we consider the flare stars in clusters: in the latter, the brighter photospheric backgrounds are such that only the largest flares are detectable, and these are systematically shifted to longer intervals.

As regards the slopes of the flare energy spectra for solar neighborhood flare stars, the mean slopes reported by G02 for M1-M5 stars, i.e. <b> = 0.75±0.19, correspond, in the notation of eq. (3), to s = 1.33±0.36. For the Kepler flares on M dwarfs, Davenport's (2016) result <b> = 0.61±0.20 corresponds to s = 1.64±0.6. And for the ultracool Kepler dwarfs, Paudel et al's (2017) result <b> = 0.68±0.21 correspond to s = 1.47±0.4. Three of these ranges of <b> overlap (within 1σ) with the prediction (s = 1.5) of the compact loop model in eq. (10). These comparisons suggest that the concept of twisting instability in coronal loops may prove helpful in interpreting the properties of flare energy spectra in the Sun and in solar neighborhood M stars.

How can we understand the decline in stellar flare rate with increasing stellar age, i.e. with increasingly slow rotation (Davenport 2016: his Fig. 7)? We suggest that an argument akin to that at the end of Section 12 above may be applicable. Slower rotation means systematically weaker dynamo field strengths: see (e.g.) Pipin and Kosovichev (2016, their Fig. 5). Therefore, older stars will generate field strengths which are weaker than similar stars of younger age. As a result, eq. (10) indicates that the flare energies will be reduced in general. As a result, more of the flares in an old star will fall into regime (ii), thereby becoming "lost in the noise". However, the flares which are large enough to rise above the noise, and find themselves in regime (iii), will still obey eq. (10), and will therefore still exhibit a power law index of 1.5.

### 14. Slopes of Flare Spectra in the Coolest Dwarf Stars

In our calculation of flare frequency in terms of a loop being driven unstable, we assume that a granule with horizontal diameter λ will drag field lines with step size λ in a random walk across the solar surface. But in order for this mechanism to operate effectively, this dragging of the field requires that the field be "frozen in" to the convecting gas. This condition will be satisfied if the electrical conductivity $\sigma_e$ is large enough. The fact that we have been able to apply the hypothesis of random walk to driving a coronal loop unstable in the Sun suggests that in the solar atmosphere, the frozen flux condition works quite well. It may also be valid for M dwarfs at least as late as M6.5e (the latest type in the G02 sample for which coefficients a and b (see eq. (1)) are given.

However, at some point, when we consider the coolest stars, where $\sigma_e$ values are expected to decrease significantly compared to the value in the Sun, the frozen flux assumption will break down. In such stars, magnetic fields will move from their original position not only because flowing gas is pushing them, but also by means of diffusion. To quantify this discussion, we note that in a magnetic flux tube of linear extent D, the time-scale on which significant diffusion occurs is of order

$$t_d = \sigma_e D^2 /\pi c^2 \qquad (12)$$

where $\sigma_e$ is in electrostatic units, and c is the speed of light (Bray and Loughhead 1964: hereafter BL64). Thus, in the course of the time interval T between flares, the magnetic fields in a coronal loop will not only be jostled randomly by granules, but will also diffuse away from their starting location by a distance

$$D = c \sqrt{(\pi T/\sigma_e)} \qquad (13)$$

How large is D in the case of the Sun compared to the size of a granule? To answer this, we need to know the photospheric values of $\sigma_e$: these are $7 \times 10^{11}$ e.s.u. (BL64). Let us consider the case of flare intervals T = 1-10 days: such intervals are in the middle of regime (iii) as plotted by G02 for the Sun. Then as the Sun "waits" to built up a particular loop to instability, the fields in the loop will diffuse away from their initial locations by distances of D ≈ 100-300 km in the photosphere. These distances are smaller than the typical granule size (λ = $10^3$ km) by factors of up to 10: as a result, field lines which started inside a certain granule will be (more or less) still contained within that granule by the time T at which instability is reached. Therefore it is reasonable for us to consider granular jostling (due to frozen-in magnetic fields) as the dominant process which operates on such coronal loops on the Sun.

But as we go to cooler stars, the value of $\sigma_e$ decreases, thereby increasing the value of D. Moreover, λ is also decreasing in cool stars (see Section 14). Therefore, we expect that in the coolest stars, the process of moving magnetic field lines (thereby building up to loop instability), will be affected not only by a random walk on scales of λ, but also by diffusion on scales of D. In cases where D exceeds λ, eq. (5) may be altered to take on the form T = (L/D)$^2$ τ. Inserting D ~ √T, this leads to $T^2$ ~ $L^2$, i.e. T ~ L. Combining this with eq. (7) for the smaller flares, we find that in regime (iii), the flare energy E on such stars should scale with T according to

$$E \sim T^3 \qquad (14)$$

This is a steeper dependence than we found above for regime (iii) flares in the Sun. In terms of eq. (1), the b value corresponding to eq. (14) will be b = 0.33. Such a value is definitely smaller than the b values which occur in M1-M5 dwarfs, for which G02 lists <b> = 0.75.

For larger flares, where the energy E is given by eq. (8), we expect $E \sim T^2$, corresponding to b = 0.5.

The small value of b = 0.33 may be helpful in interpreting a result reported by Paudel et al (2017) concerning the energy spectrum of flares on an L0 dwarf. In one of their L0 dwarfs, they obtained a value of b = 0.34±0.04. We hypothesize that in this star, effects of low conductivity may be contributing to the value of b.

The transition from random walk to diffusion is expected to occur in stars where D becomes comparable to, or larger than, λ. Assuming that λ is related to the pressure scale height, λ will scale as T(eff)/μg. The variables T(eff), μ, and g all vary as we go down the main sequence, but only by small factors (of order 2 or so). And D will scale as √(T/$\sigma_e$). Here, √T will change only slightly as we go down the main sequence. The major change in the ratio D/λ will be due to the value of $\sigma_e$. In a partially ionized gas, $\sigma_e$ is given by

$\sigma_e = 10^{17} (p_e/p)(1/\sqrt{T(eff)})$     e.s.u.     (15)

(see BL64, assuming a collision cross-section of $10^{-15}$ cm$^2$ ). In the solar photosphere, where T(eff) = 5777 K, models indicate that the value of $p_e/p$ is 5 x $10^{-4}$ (Vernazza et al 1981), leading to $\sigma_e$ = 7 x $10^{11}$ e.s.u., as mentioned above. The electrons in the solar photosphere are contributed by metals, mainly Na, Mg, Al, Si (with ionization potentials I.P. = 5-8 eV). If we could arrange to have an increase in D by a factor of 10 or more, this would make D comfortably larger than λ. One way to achieve this is to have $\sigma_e$ in the coolest decrease by a factor of 100 or more compared to the Sun. This would require $p_e/p$ to decrease to a value of ≤ 5 x $10^{-6}$. What could cause such a reduction in electron density? The progressive decrease in temperature as we go down the main sequence, with an accompanying decrease in the ionization of elements with I.P. = 8 (Si), 7 (Mg), 6 (Al), will have this effect. At low enough temperatures, only the alkali metals (K, Rb, Cs, with I.P. ≈ 4 eV) will supply electrons: the cosmic abundance of K is smaller than Mg or Si by factors of a few hundred, and this could account for a reduction in $\sigma_e$ by a factor of 100 or more. It would be interesting to use model atmospheres to determine at what values of T(eff) the ratio of $p_e/p$ falls to a value as low as ≈5 x $10^{-6}$ in the photosphere. In this regard, MacDonald (2017) has examined the ratio of $p_e/p$ in his model of a star with T(eff) = 2900 K, corresponding to spectral types M5.5-M7.5 (according to results of Rajpurohit et al 2013): MacDonald finds that in the photosphere of such a star, $p_e/p$ has values of order 5 x $10^{-7}$. And in the photosphere of an L0 star, with T(eff) = 2250 K), he finds $p_e/p$ ≈ 2 x $10^{-8}$. Such values easily satisfy the criterion that is required to make D larger than λ mentioned above: $p_e/p$ ≤ 5 x $10^{-6}$.

In view of this result, we suggest that M dwarfs with spectral types of M5 or earlier should exhibit $E \sim T^s$ behavior where s is close to the value (s=1.5) we found for solar-like flares, whereas in dwarfs of spectral type M5.5-M7.5 or later, the value of s can take on values that are as large as s = 3.

In a completely independent study of magneto-convective models of low mass stars, MacDonald and Mullan (2017) have found that in order to replicate the observed field strengths at the surface of two flare stars with spectral types M5.5 and M6, allowance must be made for the effects of finite electrical conductivity. We consider it noteworthy that effects of finite conductivity have shown up in two independent studies of physical conditions in late-M dwarfs.

## 15. Discussion

In this section, we discuss some general aspects of flares in the context of the work we described above.

### *15.1. Systems for classifying flares*

In this paper, we have considered a flare classification which assigns flares to one of three regimes based on the flare energy. Other authors use different systems to group flares into different classes. For example, Hawley et al (2014) consider flares as belong to two groups: classical and complex. This is a useful classification scheme when one considers the number of peaks in a flare light curve: a flare with a single peak in its light curve is considered to be "classical", whereas a light curve with multiple peaks is classified as "a complex flare". Hawley et al demonstrate that the energy in a complex flare on a given star generally is larger than the energy in a classical flare on that star.

Since, in this paper, we have assigned flares with large energies to regime (i), and flares with smaller energies to regime (iii), we might ask: do our regimes (i) and (iii) correspond to complex and classical flares (respectively) in the notation of Hawley et al (2014)?

In order to address this question, we note that the essential assumption in the present paper is that all of the flares which we consider involve only a single coronal loop. And that simple assumption applies not only to regime (iii) flares, but also to regime (i) flares at high energies. Both involve a single loop, but the loop in regime (iii) is 3-D (see eq. 7), whereas the loop in regime (i) is 2-D (see eq. 8). Our single-loop scenario suggests that once the loop goes unstable, there will be a single release of flare energy. After that release, there are two options. First, the loop might be completely disrupted by the flare, and the loop would lose its identity. In such a case, there would never be another chance to go unstable. In view of this, we see no difficulty with identifying the classical flare of Hawley et al (2014) with the type of flare we consider in the present paper in either regime (i) or regime (iii).

Second, after a flare occurs in the way we describe in this paper, the loop may (in some circumstances) not undergo complete disruption, but may instead survive as a loop. If that happens, then after a long time of 500 min-$10^5$ min (according to our estimates in Section 10), the loop might be driven unstable once again by granule flows and undergo another flare. Might such an event qualify as a "complex flare" according to Hawley et al (2014)? According to Hawley et al, the multiple peaks in a complex flare occur within tens of minutes of each other: these intervals seem too short to be associated with the time required for a loop to undergo a second incident of instability in the presence of the granule-dragging considered in the present paper. In this sense, we consider it unlikely that the complex flares described by Hawley et al (2014) could be equated to the high energy (2-D) flares which we assign to regime (i).

In our opinion, the complex flares discussed by Hawley et al (2014) may belong to a different category from the single loop flares discussed in the present paper. As already mentioned above in Section 5, solar flares come in two principal categories: compact flares and 2-ribbon flares. In the latter, more than one loop is involved in the flare process, so naturally a more complicated light curve is expected to occur. Perhaps complex flares belong to the 2-ribbon class. Such a process is not covered by the single loop treatment in the present paper.

Another option for interpreting the complex flares described by Hawley et al (2014) is based on the presence of time-scales of tens of minutes between multiple peaks in a complex flare. Based on those time-scales, it can be argued (on the basis of MHD wave properties) that sympathetic flaring might be responsible for the second (and later) peaks (Gizis et al 2017). Such a process is not covered by the mechanism we discuss in the present paper.

### *15.2. Is a broken power-law appropriate for all flare star spectra?*

In our discussion, we have referred to flare energy spectra as containing two distinct power law regimes (i) and (iii), i.e. a broken power law. We have suggested 2-D versus 3-D models as a possible interpretation of the broken power law: a steep slope at lower energies, and a shallower slope at higher energies. Our model suggests that the break in the spectrum should occur at a critical energy E(c) when the size of the flaring loop exceeds a local scale-height. The value of E(c) will depend on local field strengths and local densities: E(c) is not expected to be the same in all stars. There is no physical reason to expect, *a priori*, that in any particular star, the value of E(c) necessarily falls inside the range of flare energies which are actually detectable from that star. This suggests that a spectral break *need not* be a feature of *every* empirical flare spectrum. Observational selection allows flares to be detected over only a finite range of energies for any particular star, and that range might not contain E(c). In fact, not even all of the flare stars which we selected to illustrate in Figure 1 (above) show clear indications of a spectral break.

Finite ranges for flare energies are also inevitably present in the Kepler data reported by Hawley et al (2014): the most active star has flare energies spanning a range of 2-3 orders of magnitude, while the least active star has a flare energy range of only 0.5 orders of magnitude. In the Hawley et al sample, only the star with the broadest range of energies (GJ1243) shows a spectral break at the higher energies.

The class of stars for which the widest range of flare energies is available is the class of solar-like stars: extensive data have been available for the Sun for decades, and the Kepler spacecraft was targeted especially on observing solar-like stars. With such a broad range of energies to study, we might hope to have a better chance of having E(c) fall somewhere inside the observed range of energies for flares on solar-type stars. As a result, we might hope to identify a spectral break such as the one we predict in this paper, i.e. shallower at high energies, and steeper at lower energies. Kepler data on the largest events ("superflares", with E = $10^{33-36}$ ergs) have been analyzed by Shibayama et al (2013): in terms of the differential energy spectra (see Section 2 above), they obtained values of the index α of 2.2 (the most active stars), and 2.0 (the least active stars). Converting to the integral index b = α-1, and then to the E-vs-T index s = 1/b (Section 4), we find that the Kepler superflares reported by Shibayama et al have s = 0.83-1.0. This range overlaps with the value s = 1 we predict for regime (i) flares (see eq. 11 above). Therefore, the superflares have a power law index which is consistent with our scaling for the largest (i.e. 2-D) flares in granule-driven loops.

What about lower energy flares, i.e. the flares we would assign to regime (iii)? Shibayama et al (2013) supplement their Kepler superflare analysis with spectra obtained by previous researchers for smaller solar flares: nanoflares, microflares, and "solar flares", with energies of $10^{24-26}$ ergs, $10^{27-29}$ ergs, and $10^{27-32}$ ergs respectively. The differential power law spectra for these three groups are listed by Shibayama et al: they correspond to spectral indices s = 1.3, 1.35, and 1.89. These indices for lower energy events are definitely steeper than the index for the superflares (s = 0.83-1.0). In fact, the slopes for the lower energy events overlap with the prediction s = 1.5 which our model predicts for flares in regime (iii), i.e. for the smaller (3-D) flares in granule-driven loops. If our interpretation is valid, we suggest that for solar-like stars, the critical energy E(c) lies between $10^{32}$ ergs (the largest flare energy in the solar data) and $10^{33}$ ergs (the smallest flare energy in the Kepler superflare data).

### *15.3. What can be learned from empirical slopes which differ from 1.5 and 1.0?*

In our derivation of the power laws 1.5 and 1.0 for 3-D and 2-D flares respectively, we assumed that there are no systematic variations in $\lambda$, $\tau$, $z$, and $\Delta H^2$ from one flare to the next on any particular star. Are these assumptions plausible?

As regards the granule size $\lambda$, this quantity is determined mainly by the stellar gravity and the surface temperature. Magnetic effects might possibly be expected to interfere with convective properties if a flare location were to approach too closely to a sunspot or a pore: however, an empirical study of granule diameters at varying distances from a sunspot pore (Darvann and Kusoffsky 1989) reveal no significant change in $\lambda$ with distance.

As regards the granule lifetime $\tau$, this is observed to increase as one approaches a pore (Darvann and Kusoffsky 1989): the lifetime of an average granule is about 5 minutes far from the pore, but increases to 12 minutes close to the pore. However, we have already (see Section 8) assumed that $\tau$ has values ranging from 5 to 10 minutes. Thus, our treatment already allows for essentially all of the systematic increase in $\tau$ close to a spot.

As regards $z$, the magnetic scale height, the numerical value that applies to any given flare depends on how deeply below the surface the surface dipole for the local active region is actually buried (see Section 11). One might expect that in any given star, an active region with weak fields would correspond to a dipole which lies far below the surface. In such a case, the loops emerging into the corona would be large: this would lead to larger $z$, in fact with values becoming almost as large as the stellar radius. On the other hand, if the fields are strong, then the equivalent dipole would lie closer to the surface. In such a case, the coronal loops would be typically smaller, leading to a smaller value of $z$. This suggests that there could be a systematic increase in $z$ as L increases: we represent this as $z \sim L^x$ where x is positive.

As regards $\Delta H^2$, the argument in the previous paragraph suggests that magnetic field strengths in a particular star probably scale inversely with L. To represent this, we guess $\Delta H^2 \sim L^{-y}$ where y is positive.

To the extent that these scalings are pertinent, we can return to eqs. (7)-(9) and re-evaluate the slopes that we expect to occur in the flare spectra. No changes are necessary for eq. (9) since $\lambda$ and $\tau$ do not vary systematically. But eq. (7) changes to $E(3) \sim L^{3-y}$. As a result, E(3) scales now as $T^s$ where  s = 1.5 - 0.5y: thus, regime (iii) flares are predicted to have spectra which are not as steep as we previously derived (1.5). As regards E(2), eq. (8) to $E(2) \sim L^{2+x-y}$. As a result, E(2) scales now as $T^s$ where  s = 1 + 0.5(x-y). Thus, depending on the algebraic sign of x-y, the spectra of regime (i) flares are predicted to either steepen or become even shallower.

In studies of flare spectra, although we have arrived (in the simplest cases) at two preferred values for the slopes (1.5 and 1.0 for regimes (iii) and (i)), the discovery of departures from these preferred slopes in any particular star may contain information about the indices x and y defined above. For example, inspection of Figure 1 shows that the star Ple1 has a spectrum that is shallower than the preferred 1.0 slope for regime (i) flares: this suggests that x-y is negative in this star. On the other hand, TC 275 has a spectrum that is steeper than 1.5, suggesting that we are observing regime (i) flares in a situation where x-y takes on a positive value.

### 16. Conclusion

In this paper, we have examined the following hypothesis. When field lines are frozen into the photospheric gas at the foot-points of a coronal loop, the field lines are subject to random walk by "jostling" due to convective (granular) flows. In the presence of this random walk, the loop can, in certain conditions, be driven unstable in a way which is relevant to flaring activity. We find that this hypothesis leads to results which are helpful in interpreting observations of flares in the Sun and on stars: (i) the estimated time-scales between flares are consistent with the observed frequency of flares; (ii) the observed energies in flares can be obtained from local magnetic fields if even a small fraction of the field undergoes reconnection; (iii) the energy E of relatively small flares is found to be related to the time interval T between flares according to a power law $E \sim T^s$ with index s = 1.5; (iv) the energy E of relatively large flares is found to scale as a shallower power law of T, with an index s = 1.0.

As regards item (iii), it is noteworthy that the largest sample of solar flares ever examined (containing more than 56,000 X-ray flares, spread over 30 years) is found to have a power law energy spectrum with s = 1.50±0.01. This agrees very well with the index we predict in relatively small flares.

Moreover, in the largest sample of stellar optical flares ever examined (the Kepler sample, with almost one million flares on 4000+ flare stars), the value of s for the subset of M dwarfs is found to be 1.64±0.6: this also overlaps with the value s = 1.5 we predict in relatively small flares. And in the largest stellar flares (superflares in solar-like stars), the spectral index is found to be s = 0.83-1.0: this range overlaps with the index we predict for relatively large flares. However, whether we refer to "relatively small" or "relatively large" flares, it is important to emphasize that *all* of the flares we discuss in our model are assumed to occur in a single coronal loop. We make no attempt to predict the properties of other classes of flares, such as two-ribbon flares or sympathetic flares.

As regards the assumption of frozen-in field lines, we expect that in the coolest stars, lower electrical conductivity will eventually lead to a breakdown of the frozen-flux assumption. In such cases, we predict that the index s will increase to a value of 3.0 in intermediate flares (regime (iii) in the notation of Section 1 above), and to a value of 2.0 in large flares (regime (i)). That is, the regime (i) flares will still exhibit shallower spectra than the regime (iii) flares. The value s = 3 in regime (iii) is consistent with observations of flares on an L0 dwarf reported by Paudel et al (2017).

**Acknowledgements**

We appreciate the constructive criticisms of an anonymous referee. We thank Dr J. MacDonald for providing information about conductivity in cool star photospheres. DJM appreciates greatly the assistance given by Dr W. H. Matthaeus in preparing Figures 2 and 3. DJM is grateful to God for the original idea.

**References**

Alfven, H. and Carlqvist, P. 1967, Solar Phys. 1, 220 (AC)

Aschwanden, M. 2016, ApJ 831, 105

Babcock, H. W. 1961, ApJ 133, 572

Bray, R. J., Cram, L. E., Durrant, C. J.. & Loughhead, R. E., 1991, Plasma Loops in the Solar Corona, Cambridge University Press, pp. 125-127

Bray, R. J. and Loughhead, R. E. 1964, Sunspots, NY: Dover, pp. 123, 124, 270

Darvann, T. A. & Kusoffsky, U. 1989, in Solar and Stellar Granulation, eds. R. Rutten and G. Severino, Kluwer Academic Publ., p. 313

Davenport, J. R. A. 2016, ApJ 829, 23

Davenport, J. R. A., Hebb, L., & Hawley, S. L. 2015, ApJ 806, 212

Drake, J. F. 1971, Solar Phys. 16, 152

Eddy, J. A. 1979, A New Sun: the Solar Results from Skylab, NASA SP-402.

Gershberg, R. E. 2002, Solar Type Activity in Main Sequence Stars, Astroprint, Odessa, Russia (G02)

Gizis, J. E., Paudel, R. R., Mullan, D. J. et al 2017, ApJ 845, 33.

Hawley, S. L., Davenport, J. R. A., Kowalski, A. F. et al 2014, ApJ 797, 121

Hood, A. W. and Priest, E. R. 1979, Solar Phys. 64, 303 (HP)

Kasinsky, V. V. and Sotnicova, R. T. 2003, Astron. Astrophys. Trans. 22, 325 (KS03)

Kittinaradorn, R., Ruffolo, D. & Matthaeus, W. H. 2009, ApJ Lett 702, L138 (KRM)

Knizhnik, K. J., Antiochos, S. K., DeVore, C. R., & Wyper, P. F. 2017, arxiv 1711.00166

Kuhar, M., Krucker, S., Oliveros, J. C. M., et al. 2016, ApJ 816, 6.

MacDonald, J. 2017, personal communication.

MacDonald J. and Mullan, D. J. 2017, arXiv 1711.09434

Mullan, D. J. 1976, ApJ 207, 289

Mullan, D. J. 1979, ApJ 231, 152

Mullan, D. J. 1984, ApJ 282, 603

Mullan, D. J. 2009, Physics of the Sun: A First Course [CRC press], pp. 77-81, 296, 337.

Mullan, D. J., Mathioudakis, M., Bloomfield, D. S. and Christian, D. J. 2006, ApJ Suppl. 164, 173

Paudel, R. R., Gizis, J. E., Mullan, D. J. et al. 2017, ApJ (subm.).

Pipin, V. V. & Kosovichev, A. G. 2016, ApJ 823, 133

Priest, E. R. and Milne, A. M. 1980, Solar Phys. 65, 315

Pustil'nik, L. A. 1986, Soviet AJ Lett 14, 940

Rajpurohit, A. S., Reyle, C., Allard, F. et al. 2013, A&A 556, 15

Schmelz, J. J., Holman, G. D., Brosius, J. W., & Willson, R. F. 1994, ApJ 434, 786

Shakhovskaya, N. I. 1989, Solar Phys 121, 375 (Sh89)

Shibayama, T., Maehara, H., Notsu, S. et al. 2013, ApJS 209, 5

Shulyak, D., Reiners, A., Engeln, A. et al. 2017, Nature Astronomy Lett. 1, #0184.

Spicer, D. S. and Brown, J. C. 1981, The Sun as a Star, ed. S. Jordan, NASA SP-450, p. 413.

Svestka, Z. & Howard, R. 1979, Solar Phys. 63, 297

Torres, G., Andersen, J., & Gimenez, A. 2010, Astron. Astrophys. Rev. 18, 67

Van Hoven, G. 1981, in Solar Flare Magnetohydrodynamics, ed. E. R. Priest, NY: Gordon and Breach, p. 217

Vernazza, J. E., Avrett, E. H., and Loeser, R. 1981, ApJ Suppl. 45, 635

Veronig, A., Temmer, M., Hanslmeier, A., Otruba, W. & Messerotti, M.  2002, A&A, 382, 1070

Wolfram 2000, **http://mathworld.wolfram.com/RandomWalk1-Dimensional.html**